# Towards Self-Service Governance by means of Information Technology

*Abstract*

In this paper we present a novel model for governing societies based on modern information technology, which neither relies on manual bureaucratic labor, nor depends on process-based e-government services for governance. We expose the flaws of the later and argue that it is not feasible for sustainable governance due to permanently changing laws and instead propose a model in which people can govern themselves in a self-service manner by relying on constellations of data stored in a network of governmental databases to which citizen and officials have read- and write access under rules defined by temporary valid law.

## 1 Introduction

In "The Necessity of eGovernment", Fenwick *et al.* (2008) apply Coase's theorem of transaction costs and Dahlman's theory of social transaction costs to rigorously argue that traditional government, orchestrated by humans, is unsustainable and technologically deprecated. Their basic observation is that "*governance becomes increasingly complex in a society with a substantial number of daily transactions*", whereby the burden on governance is disproportionally higher than the corresponding number of transactions.

Increased computerization of the society has resulted in a significantly increased speed and quantity of transactions, which are subject to governance; increased political self-awareness and a more and more transparent governing on the other hand has additionally increased the need for fast and accurate communications between those who govern and those who are being governed. In order to meet the demands of these new expectations from the governed – the subjects in a society, the government needs methods that would allow virtually real-time responding to the requests.

Fenwick *et al.* stress the importance of *proper* use of information technology (IT) in governance, as only appropriately automated governance can meet the increasingly complex requirements of the modern society. Only proper computerization they argue, can remove redundancies in processing and storing data needed for governance, significantly reduce costs of information collection, analysis and distribution; minimize corruption and make governing actions transparent, thus reducing the public fear of a "big brother".

The general goal of our research is to find ways how to circumvent the imperfect and potentially corrupt human factor in all aspects of governance. Our claim is that it is possible to establish a distributed information system, which would allow subjects to manage most of their relations themselves – in a "self-service" manner, being constrained only by the technical rules of such information system, which' behavior would be non-discriminative and equal to every request. We claim that such a system would liberate governments from the need to themselves actively collect and analyze data required for governance, resulting in a significant reduction of transaction costs.

We shall use the design-science research methodology (Hevner et al. 2004) to search for (and describe) a model for an information system that would allow self-service governance according to hereinabove defined hypothesis. Rather than relying on empiric research, design-science focuses on the *design* of novel artifacts that are created to solve heretofore unsolved problems. The validity of research is thus grounded on both the relevance of the problem and the feasibility of the offered solution.

In this article we describe the general architecture and specific characteristics of self-service governance (ss-gov), further we describe its stakeholders and their roles. The outline of the article is as follows: in the introduction we define the theoretical framework for our research and emphasize the unsustainability of e-government as the main competitor of our model. In chapter 2 and its subchapters we describe the concept and vision behind ss-gov as a novel model for governance based on constellations of legal relations; thus, in chapter 2.1 we describe the idea behind ss-gov based on a realistic scenario, and in chapter 2.2 we describe the conceptual architecture of the ss-gov model and its stakeholders, while drawing analogies between governance of real-world societies compared to virtual communities in online games. Finally, in chapter 3 we will try to evaluate the feasibility of ss-gov against two common real-world scenarios.

### 1.1 Theoretical Framework

Human society is a dynamic system of subjects, which is controlled by a certain authority and organized according to specific rules. The society as a system behaves chaotically, and it is impossible to predict its state in the future despite available knowledge on its state in the past. Nonetheless, certain laws of nature exist in accordance to which this system behaves, which are subject to research of social- and legal sciences.

Thus, political philosophy teaches us that the beginning of a society is the *social contract* – an implicit agreement between the members of a community, which regulates how the society will be governed. Social contract theory (Hobbes 1651; Locke 1689; Rousseau 1762) – which is yet today one of the most influential modern theories (Friend 2004), teaches us that the society is a system that consists of (i) *subjects*, which are governed by (ii) the *sovereign*, as well as (iii) *rights*, which the sovereign – as the origin of all rights, grants to its subjects (Rousseau 1762, bk. 1, ch.9).

Also modern legal theory uses this knowledge from social contract theory (Jellinek 1905, 32; cf. Boyle 1993) to model the relations among the members of a society. In jurisprudence, every subject (natural person, company, club, state, etc.) has *legal subjectivity*, which means that it can have (legal) rights and be subject to duties (Cerar 1996, note 4). Legal subjectivity is granted by the sovereign to people or associations (Jellinek 1905, 77) in form of a *legal status* – an attribute that allows subjects to get rights (Jellinek 1905, 78–79; Bauböck 2010). *Rights* (as well as all legal relations) in mainstream legal theory – as defined by Hohfeld (1923), are relations between exactly two subjects (Hohfeld 1923; Lazarev 2005; cf. H.E. Smith 2011) whereby one direction of such relation defines the entitlement (to claim protection by the sovereign (Jellinek 1905, 77) while the other defines the corresponding duty (cf. Vodinelić 1976 in: Cerar 1996, 10).

In modern societies, the sovereign is the democratic state, which acts trough its institutions (cf. Kersten 2000, 253–61). The sovereign is the source and protector of all legal relations as well as the origin of citizenship and other legal statuses. The management of those and the protection of their integrity is therefore the topmost priority of the sovereign – it is an essential task of governing.

Based on these premises we may abstract society to being a network of subjects, which are bound to each other trough Hohfeldian legal relations, the so-called "bundles of sticks" – or "bundles of rights" (Henry E. Smith 2004 esp. note 30; D. R. Johnson 2007), whereby the crucial task of governance is the protection of the integrity of this *network of legal relations*.

Traditionally, state officials of varying ranks are governing this network of legal relations, constituting the service of *public administration* – PA. Human-based PA was until recently the only plausible way to manage complex reasoning needed for governing human relations – simply because no other way existed to intelligently capture, communicate, store and interpret information required for governance.

A permanent issue with human-based PA is corruption (cf. G. E. Caiden and Caiden 1977; S. Johnson, Kaufmann, and Zoido 1999; Köchli 2006; Noor 2009; Pinterič 2009, 206–14), which influences the accessibility, timeliness, quality and cost of public services. If we see the society as a system managed trough PA, then every human inside PA presents a potentially significant security hole, which can be exploited to compromise the system. But even if traditional PA would be completely free of corruption, important issues such as slow responsiveness (Vigoda 2000; Paulin 2010; Popescu 2011) will remain. And last but not least, there remains the undisputed truth of Parkinson's laws, which are best known for their proof that "*work expands so as to fill the time available for its completion*" (Parkinson 1955).

### 1.2 Legal Certainty in Modern E-Government

Fenwick *et al.* (2008) see as the solution to the rigorously described problems the use of e-government. The research community vastly agrees that the focus of e-government is on *services*, which the government offers to its citizens (Marche and McNiven 2003, 75; Al-Sebie and Irani 2003). Such services can be applications to search for employment (cf. Celino et al. 2010), public information catalogues (cf. Veljković, Bogdanović-Dinić, and Stoimenov 2011), municipal Twitter-feeds (cf. Mambrey and Dörr 2011), government registries (cf. Lenarčič 2009), legal information systems (cf. Lesjak and Jagodnik 2009), etc. E-government services would be according to modern definitions also governmental surveillance tools like e.g. the German *Bundestrojaner*[1] (cf. Berlit and Wegewitz 2008).

"Services" by the definition of the word imply a process-oriented approach towards *delivering* them – the service in the restaurant is carried out by the waiter, while taking food from the buffet is not service, but self-service. This process-oriented approach of e-government however bears several severe flaws that have not yet been appropriately discussed: modern e-government lacks to answer important issues such as the issue of *sustainability* of services, their compliance with the principle of legality (ergo *legal certainty*) and their *agility* (Paulin 2011a). (A further flaw, namely *openness*, has been an issue already prominently emphasized by Tim O'Reilly (2011) in his article "Government as a Platform", where he uses Kettl's metaphor of the vending machine (Kettl 2008) that offers a limited set of overpriced items and compares it to the blossoming market-place of Raymond's (1999) bazaar or its digital alternatives, like iPhone's market for application, which allows commercial providers to use the phone's infrastructure for delivering new and better services.)

---

[1] lit.: federal Trojan horse
[2] See (Mayer-Schönberger 2008) for a critical response.
[3] ISO/IEC standard 7498-1:1994
[4] The *Hypertext Transfer Protocol* is a de-facto standard protocol for the communication with Web servers.
[5] The *Hypertext Markup Language* is nowadays just one of many standards for digital

Legal theory distinguishes between public law and private law, whereat public law regulates the relations between the state and its citizens – ergo relations between the sovereign and its subjects, while private law regulates relations amid the subjects based on their will (cf. Toplak 2008, 23; Horwitz 1982). A major difference between public and private law is, that private law restricts the freedom of the subjects, while public law empowers the sovereign.

In a society that adheres to the rule of law, the sovereign (ergo the state trough its bodies) operates in accordance to the principle of legality, which means that every action and every decision made by the state must be explicitly defined by law. It is important to be aware that this applies both to stated decisions, as well as the procedures that lead to such decision (cf. Jerovšek 2000, 28–29). This elementary legal principle allows subjects not only to exercise control over the sovereign, but guarantees also legal certainty, which makes the sovereign's actions transparent and foreseeable. Legal certainty prevents the state's bodies (e.g. government, police, judges, etc.) to act or decide arbitrarily, which is crucial, as state arbitrariness would break crucial legal principles, such as the equality before law (Šinkovec 1998, 31).

With the advent of e-gov applications, we must ask ourselves how to apply the principle of legality to this new dimension. From a technical perspective, services offered trough e-gov are often procedures that the subject can trigger (and communicate with) remotely, whereby such procedures are executed on the sovereign's server. The communication between the subject and the sovereign is in fact the technical communication between the terminal equipment of the subject and the serving terminal of the sovereign.

Unlike human-to-human communication, which bases on the interpretation of analogue messages, the digital communication is discrete, exact and unambiguous. Human communication for example does not rely on strict grammar or correct pronunciation of words – two foreigners will be able to perfectly communicate in English despite their ignorance of its formal rules. Also general human perception is based on the interpretation of analogue, ambiguous information – we can visually recognize an object being a car even though we have never seen it in the exactly same environment, angle, shape, etc. However, it is impossible for two computers to communicate without adhering to strict protocols that regulate the exact semantics of the transmitted signals and information. We can say, that human interaction is analogue and computer interaction is digital – the former is ambiguous and therefore must be interpreted while the later is by nature unambiguous.

Because of this digital nature of electronic interaction, it is also e-government that must obey the rules of the digital world. This brings us back to the problem of legal certainty: Because in the real – analogue – world the interactions are subject to interpretation, legal certainty can be achieved by defining spaces of freedom within which the interaction must take place. Thus for example, the administrative procedure is initiated upon submitting a written and signed application (URL 2006, vol. 24, para. 63) to the responsible administrative body. There is no legal definition of which material the application must be written on, the font that must be used or the color of the writing or the exact formulation – because the interaction takes place in the analogue dimension, it can be reasonably expected from the administrative body to be able to understand the content and demand of the application. If however the application would be an electronic document sent to an endpoint of an e-government system, then in that case the structure and semantics of the administrative application must have been rigorously defined.

For thousands of years law has been governing only the relations between human beings – through limiting the freedom of its subjects, the sovereign regulated their behavior. Lessig (1999, 506–507) describes four constraints to regulate[2]: laws, social norms, the market and architecture. Law regulates by ordering people to behave in a certain way and if they do not, it threatens them with punishment – e.g. we may drive our car on the highway only within a certain speed limit, if we exceed it, we risk to get fined. Also social norms regulate trough threat of punishment – e.g. social isolation, but unlike laws they are not imposed "top-down" by the authority (sovereign), instead they are enforced by the community itself. Markets regulate trough the offer/demand ratio and the sovereign can influence them by e.g. forbidding imports or imposing taxes.

The fourth modality is architecture, which regulates by restricting the physical environment. In the real world, these are streets, which divide neighborhoods, bridges that connect shores or squares where people meet. Regulation trough architecture was used e.g. to prevent uprisings in Paris (Lessig 1999, note 18), corruption in Germany (Lessig 1999, 507) or child prostitution in Vienna (News.at 2008).

According to Lessig, the architecture of the cyberspace is its "code", by which he means "*the software and hardware that make cyberspace the way it is*". Lessig strictly distinguishes "code" – ergo the architectural characteristics of any information system (cf. Lessig 1999, 507–10), from law and in an earlier work he asks rhetorically (Lessig 2006, 323): "*if code is law, then obviously the question we should ask is: Who are the lawmakers? Who writes this law that regulates us? What role do we have in defining this regulation? What right do we have to know of the regulation? And how might we intervene to check it?*"

Lessig's questions are justified, however they remain unresolved. As his work focuses on intellectual property and privacy on the Web, he does not deal with the legality of public "code". Nonetheless he mentions the FBI's e-mail surveillance system "Carnivore", for which he demands that its code should be transparent (Lessig 2006, 141). Also the broader research community has so far not yet dealt with legality aspects of e-government services' architecture: Biegel (2003), Lastowka & Hunter (2004) for example discuss the authorities' tendencies to regulate cyberspace as such, while Lundblad (2007) is interested in the question whether governments are allowed to hide government data from search engines.

---

[2] See (Mayer-Schönberger 2008) for a critical response.

Lessig's understanding of the cyberspace as code is only partly correct. Wrongly he mixes all layers of the *Open Systems Interconnection* (OSI) reference model[3] with the business logic, databases and interfaces of applications into one big entity, which he names "code", while it would be right to separate them and treat them individually. As "code" – ergo *architecture* of cyberspace, only the layers of the OSI model could qualify, as it describes the complete infrastructure (starting virtually with bits over cupper cable) of an information system in terms of potentially standardizable communication protocols.

Electronic applications that can be consumed over the Internet or Web on the other hand are logically separate information systems that reside on top of the OSI model, which they utilize only to communicate with different systems – e.g. to receive requests and issue responses to the client.

If a user (the client) interacts with an e-gov service, then this interaction is an exchange of electronic messages, which accords to a particular schema that the serving endpoint (the server) understands. An example interaction with a Web-based e-gov service would look as follows: the user first sends a HTTP[4] message (request) to the web server of the e-government application; the server responds by sending her the HTML[5] Web page, which the user's terminal equipment may (or may not!) visualize and present as a user interface for further interaction. Trough a series of such requests, the user *consumes* the e-gov service, whereby the server has neither influence on how the client treats the response, nor does it know whether the response was received at all – the nature of the HTTP protocol follows the pattern "request-response-disconnect" (cf. Gutzmann 2001); it is therefore the responsibility of the server-side business logic to protect the integrity of the application with no regard to the input received.

The exchange of messages – the user's requests and the server's responses, in the e-government interaction is not architecture any more – it is an administrative procedure, initiated by the citizen with the goal to influence her legal statuses, rights or to just receive information[6].

E-government services are replacing public officials trough automated procedures. But in the same way as the public official is obliged to follow strict formal procedures that are transparent and known in advance, also e-gov services must be rigorously defined in a transparent way. However, while in the analogue world it is sufficient to define just the goals of the procedure, the digital world due to its architecture requires an unambiguous definition of not only the procedure, but also the location of the service on the network.

---

[3] ISO/IEC standard 7498-1:1994

[4] The *Hypertext Transfer Protocol* is a de-facto standard protocol for the communication with Web servers.

[5] The *Hypertext Markup Language* is nowadays just one of many standards for digital documents. It was initially developed to semantically describe content accessible over the Web.

[6] An often overseen detail of governing is the communication with citizens – be it in terms of providing access to data under the freedom-of-information legislation (FOIA), or just by answering requests. As recent research reveals (Pinterič 2010; Paulin 2011b, 8–12), citizen requests are often being simply ignored despite strict legislation.

## 2   The Principles of Self-Service Governance

In order to explain our concept of self-service governance (ss-gov), we first need to explore both terms – "governance" and "self-service":

Governance is a term with various definitions – thus Rhodes (1996) presents six uses of this term, namely describing (i) the minimal state, (ii) corporate governance, (iii) the new public management, (iv) "good governance", (v) a socio-cybernetic system or (vi) self-organizing networks. Marche & McNiven (2003) put "governance" in contrast to "government", whereby the former is *"the way in which decisions are made"*, while the later is *"the way in which these decisions are carried out"*, Stoker (1998) however sees between both terms no difference in output, but only difference in process. All of these definitions can be seen as different viewpoints on the same "thing" – the same system of *governing* a society. We shall for the needs of this article define governance in a minimalistic way, as: the sum of all actions needed to be undertaken by the sovereign in order to fulfill its part of the social contract. This definition includes all tasks performed by the three branches of government (legislative, judicative and executive), as well as all related services provided by institutions from the public and private sector.

Self-service as defined for the purpose of this article is not technology-based self-service - TBSS (cf. Reinders, Dabholkar, and Frambach 2008), where users access automated services, such as ATMs or vending machines. Self-service, as we have in mind, is much more rudimental, such as the self-service of opening a door, planting seeds, or assembling a LEGO toy. The difference is that when using TBSS, we are forced to use a predefined – and thus limited, set of possibilities; rudimental self-service on the other hand does not limit us: we can use the parts of the LEGO toy to assemble a house instead of a ship and we can grind the seed to flour and make bread instead.

### 2.1   Constellation-Based Eligibility Evaluation in ss-Gov

SS-Gov is therefore not the use of TBSS in governance and/or government (this would be e-government), but rudimental access to those factors that constitute them. As outlined in the introduction, an important task of governance is the management of the network of legal relations between subjects in a society. In order to allow self-service governance of this network, it is therefore essential to define and clearly separate what can be governed in such a manner and what not.

New public management (NPM) distinguishes between policy-decisions (the so-called "steering") and service delivery ("rowing"), whereby rowing – ergo government, can be outsourced to the private sector[7] (Osborne and Gaebler 1992; cf. Rhodes 1996, 655; Bevir 2009). NPM however focuses so far on the macro view on governance, i.e. policy-making/norm-setting and execution, while the management of micro-relations is to our best knowledge not studied yet.

---

[7] Some authors, however, indicate that privatization of certain public services can be questionable from legitimacy perspective (Pinterič 2011, 245–46).

Osborne & Gaebler's metaphorical boat-ride is not detailed enough to provide the picture of the whole society, as it has only two stakeholders – the captain, who dictates the course and the slaves who do the rowing and who could be more efficiently replaced by modern engines. Extending their metaphor, a passenger ship shall be assumed, which not only transports its passengers overseas, but offers them also food, accommodation, security and entertainment according to the afforded travelling category. The passengers of such a ship are a distinct community bound to a specific territory (the ship). They are governed by the crew, which upholds order during the trip and provides the passengers with the contractually agreed services. The relation between the crew and the passengers is similar to the relation between the sovereign and the society: in both cases the member of the community is the subject recognizing authority and paying a certain fee for being governed and provided with services; and if the governing and service are not to the satisfaction of the subjects, they will start a mutiny and eventually take matters democratically into their own hands.

Based on this new metaphor, we may analyze what is needed for governing this community. Let us assume a family (father, mother, 13 years old child, which will turn 14 in two days) travelling second class from Slovenia to Iran (arrival in Iran's territorial waters on the $5^{th}$ day). The ship features a wellness-center with swimming pool and sauna, whereby the later is charged extra to those not travelling first class and entrance is not permitted to children younger than 14 years. In Iranian territorial waters only single-sex usage of the sauna is allowed, whereby the age limit is 20 years. In Iran a person younger than 20 years can enter the sauna only if accompanied by a close relative of the same sex.

In this example, we have a fixed community of subjects (the passengers) and a fixed government structure (the crew), but we have to deal with many dimensions: Every member of this society has a status based on several relevant personal attributes: sex, age, booked travel category and relation to other travellers. This status entitles passengers to receive service and use the facilities. One attribute of the passengers – namely their age, is subject to time, which is relevant, as the child – let's name her Eve, will become 14 and will thus become entitled to enter the sauna on the third day of voyage. When the ship will enter Iranian territorial waters, the legal frame will change, which will influence also the passenger's legal situation, while their attributes will remain the same. Thus, on days 1-2, Eve must not enter the sauna, because she is to young; on days 3-4 she can go to the sauna, as she is now 14 already; from the $5^{th}$ day on, she can go to the sauna only together with her mother, but not her father, due to the Iranian restrictions.

From this example we can see, that the "bundle of rights" of each subject is determined by the context and the legal frame. If Eve tries to enter the sauna, then whoever controls this facility must grant or deny Eve access based on the available information about her age, family relations and the information whether she has paid the entrance fee; the action of checking these is essential to governance.

The bundle of rights is not something that can be efficiently stored, but must be interpreted every single time *based* on (i) the legal frame of the political community, (ii) data about the subject and (iii) the context of the given situation. Let us define – inspired by Leibniz' appeal "*Calculemus!*"[8], the bundle of rights as *B*, the subject as *S*, the legal frame (*L*) of the political community *P* as $L_P$, and the legal context (*C*) of the particular request *R* (e.g. the request to enter the sauna) as $C_R$. For each subject we have data about the subject, $d_S$ and based on this data, we can calculate its legal status (ς) in the given legal frame:

1)    $ς_S = ς(L_P, d_S)$

The legal status determines which bundle of rights somebody can have in a given legal frame, e.g. in a given country or other form of society. Thus, the bundle of rights of somebody in a foreign country is different from their bundle of rights in her home country. We can therefore say that both the subject's status ($ς_S$) and the legal frame ($L_P$) determine the subject's bundle of rights:

2)    $B_S = B(L_P, ς_S)$

In order to determine, if somebody is permitted to action *R*, we must first find the set of rights needed to perform it. This set of rights or eligibilities (*E*) is determined by *C* – the context of the given request, which is dependent on the general legal frame ($L_P$). To calculate set *E* for a given action *R*, we perform:

3)    $E_R = E(C, R)$, where $C \subset L_P$

If we now want to find if *S* has permission (*Y*) to a specific action *R*, then we must check if the set of required rights/eligibilities to perform *R* is contained in her bundle of rights *B*:

4)    If $B_S \supset E_R$ then $Y_R$ is true, else false

Given these formulas, we can continue searching for the essence of self-service in governance. As we see, making governing decisions (based on finding $Y_R$) and thus governance as such, depends on information available to the system ($d_S$). Thus, if the available information about Eve's age would be mistakenly stored as > 20, then she could go to the sauna every day, despite the biological truth. Also, as her family travels 2nd class, she must pay every time for her sauna visit, which she would not have to do if the stored information would be different.

A change of information – legal status of a subject $ς_S$ and data about the subject $d_S$, available to the sovereign results in a different bundle of rights $B_S$. Therefore in order to reach self-service in governance we need to be able to manipulate $ς_S$ and $d_S$ directly ourselves, or more precise: we must be able to manipulate $d_S$, because the legal status ($ς_S$) is determined by the data, which is the only tangible variable independent of the legal frame ($L_P$).

Manipulation of $d_S$ is traditionally performed trough a bureaucratic apparatus: in order to change our name, address, or marital status, we have to fill in forms and wait for applications to be processed. (Compare also Klischewski & Ukena (2010) for an attempt to optimize the process using semantic technologies, albeit unrelated to our research.) The same is true if we want to receive child support, get permission to drive a car or become owners of real estate.

---

[8] »... quando orientur controversiae, non magis disputatione opus erit
inter duos philosophos, quam inter duos computistas. Sufficiet enim
calamos in manus sumere sedereque ad abacos, et sibi mutuo (accito si
placet amico) dicere: c a l c u l e m u s.« (Gerhardt 1890, 7:200)

Bureaucracy – both the traditional Weberian model, as well as its modern successors, use a multiple-process-based approach to effectively determine $Y_R$. Thus, through a series of partly autonomous processes (both customer-facing processes and internal business processes; cf. (Leben and Vintar 2003)), $ç_S$ (e.g. adulthood) and $d_S$ (e.g. name, address of residency) are gathered and $Y_R$ (e.g. eligibility to change address / name / ownership of real estate) is determined. (A modernization of that approach by means of TBSS is subject to research by e-government.)

In contrast to the process-based approach of traditional governance, our vision of ss-gov is state-, or constellation-based. We reject finding $Y_R$ trough a series of cascading processes and sub-processes, as each process takes valuable time and is vulnerable to errors. Instead, we propose the manipulation of $d_S$ based on the ad-hoc calculation of $Y_R - R$ being the permission to manipulate $d_S$, conducted in one logical step.

Constellation-based reasoning[9] can be compared to a key opening a pin-tumbler lock, where the key trough its specific shape moves the pins into the right constellation, which allows the lock to be opened. Thus, the "key" for Eve, who wants to enter the sauna, would be the appropriate constellation of the information available about her (age, sex and whether she has a valid ticket or not) in the given context, while in contrast to this, process-based finding of $Y_R$ is like the doorman in front of the nightclub deciding about whom to let inside based on his interpretation of the house rules.

SS-Gov requires both read- and write-access to $d_S$, which shall be enabled directly by using the "key". Based on this self-service manipulation with the data needed for governance, the sovereign can provide its subjects all services needed to fulfill its obligations from the social contract, while being free of the burden to provide "services" that only serve to read or write data.

### 2.2 The Architecture of ss-Gov and its Stakeholders

Let us imagine living in a society managed trough ss-gov: How does the "state" look like, if we can manage our relations by ourselves? Who makes ss-gov possible? And How?

#### 2.2.1 The Electronic Registers

The idea behind ss-gov is, that the state (the sovereign) harbors a collection of registers, which contain data about its subjects, much like data is stored today in the civil registry, cadaster, or business register. Based on these collections of data, our specific legal status and bundle of rights can be determined (see chapter 2.1), given a specific context or situation. These registers are to be defined in electronic form and must be accessible trough the Internet. The entry point to each of those collections of electronic data must comply with the following requirements:

1. The interface to the system – including the location of the interface in the network (e.g. IP address or URL/URI), must be legally defined.

---

[9] A similar technique to what we call "constellation-based reasoning" was also applied by Bob Kowalsky (e.g. 1992) in his works regarding legal reasoning; compare also (Prakken and Sartor 2002).

2. The format of incoming and outgoing messages must be legally defined.
3. The procedure, how the incoming message is handled, must be legally defined.
4. Reading and writing data must be done in an analog manner, which means that only the grammar and semantics for defining the commands for accessing data can be defined.
5. Legally significant communications between the sender and the server must be provided.
6. The user has access to the core data within legally imposed restrictions.

The requirements #1-3 ensure that everybody can know, where and how to reach the endpoint of the ss-gov system, as well as how the request will be processed. Thus, the user is not bound to use specific terminal equipment or interface to interact with the system, but can hypothetically built such a system herself. These requirements can be achieved by rigorous technical standards incorporated into the legal system of the society.

Requirement #4 can be achieved by defining an artificial language for reading or writing data, or using an existing standard. Examples of contemporary artificial languages that enable read- and write-access to data would be SQL or SPARQL/Update (Seaborne et al. 2008). In our previous work (Paulin 2011c) we describe an ss-gov prototype using SQL for both manipulating data and defining access restrictions.

Requirement #5 is essential for "mashing-up" data from different sources, which are not originally linked. Let us imagine a private-sector bank, which we trust to execute a payment to our business partner only under a certain condition – e.g. our business partner must first prove that she has transferred ownership of a real estate to our name before we pay her the purchase price. The bank could provide a web application, which would receive the information from the land registry, verify it and then conduct the transaction. The bank must be therefore able to fully trust the integrity and correctness of the information. On the other hand, the ss-gov system receiving the request must be able to be absolutely sure about the identity of the sender and the integrity of the request.

The 6$^{th}$ requirement gives the subjects inhabiting the system the maximal freedom to design their legal relations in accordance with the surrounding legal frame. This freedom gives rise to new, so-far impossible uses of the state. Thus, we could imagine high-frequency trading of real estate square meters, a market for university places, changing of names according to the daily saint (if not prohibited) and absolutely transparent, if not even liquid-democratic public spending.

### 2.2.2 Lessons From MMORPG

The computerized society has many similarities with computer games in which the player's character collects items, grooms its environment and interacts with other characters. Lastowka & Hunter (2004) extensively describe the close relation between players of MMORPGs[10] and their virtual alter egos from the legal perspective.

---

[10] »Massive multiplayer online role-playing games« are online games, where multiple players from all over the world simultanuously play in the same virtual environment.

Both the "real" society and "virtual" communities are cybernetic systems in which we can have capital, do trade, have social and legal statuses, and enjoy our rights in accordance to rules provided by the sovereign. The differences between both systems could be regarded as insignificant, if not non-existent at all. Thus it is interesting to note that in the year 2003 the size of economy of Norrath – the virtual world inside Sony's Everquest, was larger than the economy of Bulgaria; furthermore, the effective pay per hour of work was at that time 3,42 $, which was more than the pay in India or China (Lastowka and Hunter 2004, 49).

From a technical perspective, a MMORPG is an information system that resides on the game provider's server and with which players interact trough specific graphic user interfaces that render to them the experience of the game. The gaming process is a permanent exchange of electronic requests and responses between the player and the server, trough which the player manipulates with its character's legal status, rights and other data relevant for the gameplay (e.g. position and appearance). Virtual real estate and other valuables are in fact only entries in the system's database (Lastowka and Hunter 2004, 51), and also trading between characters is nothing more but changing stored information. Nonetheless, the way in which the players perceive these "bits and bytes" creates in them the feeling of genuine possession and property.

Thus, Lastowka & Hunter (2004, 45) cite Lessig's report of a "nasty and protracted battle" in the virtual world of LambdaMOO between neighbors Martha and Dank. The dispute was about Dank's dog being repeatedly poisoned and killed by the flowers Martha grew in her garden. Although it is not know whether the dispute was settled before court, Lessig reports that both parties invested in the dispute the kind of passion and righteous indignation usually reserved for real world, "across-the-fence", property disputes.

Both MMORPGs and the legal system of a society are based on abstractions of the real world (Lastowka and Hunter 2004, 42; Jellinek 1905, 21). Both systems can be seen as virtual constructs that provide a platform for managing rights[11] and ensuring (legal) protection. However in contrast to the "real world", the rules in online multiplayer games are not subject to politics, democratic decisions or revolutions, which makes it possible for the authors of those virtual worlds to hard-code the rules in an absolutistic, god-like fashion.

### 2.2.3 Stakeholders and Roles

MMORPGs and other online-societies like e.g. virtual currency systems, know two distinct stakeholders: the sovereign – which is the corporation that operates the systems, and the subject, who is the player or the user of a system. Each interaction between both roles is strictly regulated by the system's architecture, which' design is at the sole discretion of the provider and generally hidden from the public. The code of such platform is at the same time law, jurisdiction, public administration and government.

"Real world" societies on the other hand are based on rules that change, that have to be politically/democratically approved and that must be transparent in order to obey the principles of legality. These rules determine how rights are created and manipulated and they define which information is needed for such actions.

---

[11] For further reading about rights, trade and reality in virtual worlds we recommend the workds of Edward Castranova and Tom Boellstorff.

A society governed trough ss-gov, will need the following stakeholders/roles: (i) *Politicians* – people, who design (descriptive) rules how the society should be regulated. These rules can be laws or other kind of policies, i.e. NPM's "steering". In the next step, these policies must be translated in a form that can be handled electronically. (ii) *Officials* are people, charged with executing law and policies, which they do in ss-gov by centrally defining the electronic registries and by defining the rules for their read/write access. Thus, their role remains similar to their role today. As the gap between the descriptive rules and their electronic implementation is bridged by human labor, which is not unmistakable, disputes could arise regarding the correctness of the implementation of law into the digital realm. To resolve such conflicts, ss-gov relies on (iii) *judges*, who – unlike today's judges, must have profound knowledge of both law and informatics. The technical security of the ss-gov system is the responsibility of (iv) *administrators*, who however only take care for the technical integrity of the system, while they must not interfere in any way with their content and logic.

Most interaction in ss-gov takes place between simple (v) *subjects* and the ss-gov system. Subjects interact with the system in order to read, write or change information based on which governance is done. They can either manipulate with the data directly by sending commands in the respective commanding language, or they can help themselves with services provided by (vi) *service providers*, who facilitate the interaction with the data. Service providers can help on different levels: they could provide programming libraries for developers of software, web applications for graphic interaction, or even human "user interfaces", whereby the subject would interact trough a human facilitator like e.g. a notary.

## 3   Evaluation of ss-Gov

Let us evaluate ss-gov in two common real-world scenarios to validate its overall feasibility. We shall assume the following scenarios: receiving child support and requesting a driver's license, both under the conditions of Slovenian law. For both we first define the data required for the right to emerge and then describe how rights can be exercised based on the stored data.

### 3.1   Child support

Child support is the eligibility to receive periodic payments from the state under the following conditions[12]:

a) That you are either the parent of the child, or the child herself. In the later case you must be older than 18 years and live in a separate household.
b) The child's registered place of permanent residence must be in Slovenia.
c) The child must not be employed or registered as self-employed.
d) The child must not be married.
e) The child must be less than 27 years old.

---

[12] http://e-uprava.gov.si/e-uprava/dogodkiPrebivalci.euprava?zdid=1064&sid=881. The conditions stated here are partly simplified or omitted in order to enhance comprehensibility.

Let us assume a civil registry *RC*, which holds the following data fields: national identification number – *RC.nin*, age – *RC.age*, address of permanent residence – *RC.adr*, and information about the relationship status to other persons –*RC.child_of* and *RC.married_to*. Further, we will need an employment registry *RC*, which holds information about who is employing whom. To define such registries in a modern relational database we could write the following SQL statements[13]:

```
1.      CREATE TABLE rc (nin, age, adr, parent_of, married_to);
        CREATE TABLE re (boss, empl);
```

Given these two registries and the defined data fields, we can instantly calculate if somebody (@*claimant*) is eligible to receive child support (for @child) as follows:

```
2.      SELECT COUNT(*) > 0      /* "true" (1) or "false" (0) */
        FROM rc INNER JOIN re ON nin = empl
        WHERE nin = @child         /* the entitled is the child */
        AND age < 27               /* if the child is not yet 27 */
        AND child_of = @claimant   /* entitled is child of cl. */
        AND married_to IS NULL     /* if she is not married */
        AND boss IS NULL;          /* if she is not employed */
```

The "right to periodically receive child support" has the effect of transferring a certain amount of money from the state budget (based on the demand of the rightful claimant) to any bank account within a given period of time (e.g. within a month). This is a typical "key & lock" situation, where the bundle of rights needed to execute the right – i.e. the transfer of money from the budget to one's bank account, is definable by existing government data. Thus, to receive child support trough ss-gov, the claimant issues a request for money transfer and the transfer is immediately conducted, if the constellation of data meets the conditions of the right.

### 3.2 Driver's license

A person obtains a driver's license upon her request under the condition[14] that she passed the exam conducted by the respective commission, is physically and psychically able to drive a motorized vehicle and has reached the mandatory age (of, let's say, 18 years).

Unlike shown in the previous example of child support, the permission to drive a car is therefore not a result of data already existing in the system, but requires new data to be generated – namely the information, if and when the claimant passed the required exam. This information must be entered into the ss-gov system by e.g. the president of the commission and as soon as this information is entered, the claimant has permission to drive on public streets. There is consequently no separate need to issue official documents, as any eligible person (e.g. policeman, employer) with appropriate read permissions to ss-gov data could verify the driving permission when needed.

In this scenario, the most important interaction is the write access of the president of the commission. The system must therefore know, who is eligible to enter new information, which could itself be data stored in a dedicated registry.

---

[13] Data definitions intentionally omitted.

[14] (URL 2008, vol. 56, para. 147); some conditions (e.g. that the applicant shall not receive permission for a certain period of time if it was taken away from her) omitted for easier understanding.

We could further imagine an inspectorate supervising the work of the commissions. Such inspectorate could have privileges to delete data about valid exams, if it for example finds that a commission is not working correctly. The inspectorate could set-up a service that would each day in the morning query the ss-gov system to find if any new permits have been issued.

Furthermore, it would make sense if the government would establish a notification service that would query the data each day and inform the citizens about any changes in their permission. Thus for example, one day you would get an email stating that the commission has given you permission to drive and the next day the system would inform you that the inspector has cancelled your permission, as it found that the commission has been corruptive.

## 4   Conclusion

In this article we described the vision of self-service governance – a model for governing a society, which needs as little as possible human, i.e. bureaucratic, intervention. In contrast to e-government, which focuses on delivering *services* to citizen and focuses on the automation of traditional bureaucratic processes, ss-gov focuses on the direct manipulation with data used for governance trough a simple, technically standardizable interface, that would be – in contrast to modern practices of e-government, in full compliance with core legal principles such as the principle of legality.

We demonstrated that governing could be abstracted to reasoning based on constellations of data stored in a governmental data network, which entitle people to do actions and enjoy rights. Constellation-based governance is superior to service-based governance, as it is agnostic to changes in the surrounding legal context, because the dimensions of legal relations are a result of information available to the government.

By evaluating the model based on common scenarios, we found that ss-gov is conceptually feasible. However, we must note that for the evaluation we deliberatively omitted technical aspects and potential problems related to them. A technical proof-of-concept of the presented model is subject to our still ongoing research and will be presented to the research community later.